\documentclass[copyright,creativecommons]{eptcs}
\providecommand{\event}{QAPL 2017} 
\usepackage{breakurl}             
\usepackage{underscore}           

\usepackage{color}
\usepackage{amsmath}
\usepackage{amssymb}
\usepackage{graphicx}

\newcommand{\erode}{ERODE}
\newcommand{\pa}{{H}}

\title{Language-based Abstractions\\for Dynamical Systems}
\author{Andrea Vandin
\institute{IMT School for Advanced Studies Lucca\\ Lucca, Italy}
\email{andrea.vandin@imtlucca.it}
}
\def\titlerunning{Language-based Abstractions for Dynamical Systems}
\def\authorrunning{Andrea Vandin}
\begin{document}
\maketitle

\begin{abstract}
Ordinary differential equations (ODEs) are the primary means to modelling dynamical systems in many natural and engineering sciences.
The number of equations required to describe a system with high heterogeneity limits our capability of effectively performing analyses. This has motivated a large body of research, across many disciplines, into abstraction techniques that provide smaller ODE systems while preserving the original dynamics in some appropriate sense.
In this paper we give an overview of a recently proposed computer-science perspective  to this problem, where ODE reduction is recast to finding an appropriate equivalence relation over ODE variables, akin to classical models of computation based on labelled transition systems.
\end{abstract}

\section{Introduction}
Ordinary differential equations (ODEs) are the primary means to modelling dynamical systems in a wide range of natural and engineering sciences. When the complexity of the considered system is high, the number of equations required limits our capability of performing effective analyses. This has motivated a large body of research, across many disciplines, into abstraction techniques that provide smaller ODE systems preserving the original dynamics in some appropriate sense (e.g.,~\cite{1098900,Iwasa1987287,okino1998,ccdtt-sci-reports-2012}). 
In particular, we refer to a number of techniques whose basic idea is to build a smaller ODE system with less variables and/or parameters, while preserving most of the original dynamics.
Depending on the specific technique used, the obtained reduced model may be used, e.g., to provide a more compact description that abstracts from low-level details, or to improve the performance and scalability of required analyses. 
The development of abstraction techniques is widely investigated in many disciplines in the science and engineering domain that deal with dynamical systems, for instance:

\begin{itemize}
\item
Ecological systems typically regard massive populations of entities that interact with the environment, and adapt to it. For this reason, quantitative abstractions have been widely studied in ecology (e.g.,~\cite{Iwasa1987287}). In this domain, abstraction techniques mainly exploit the fact that entities can range from molecules to whole organisms in their ecosystem (hence they act at different time scales), and the fact that there is often a high heterogeneity among the entities, e.g., they can have different age and metabolism, or can be located in different points in space. 
In particular, many techniques focus on the aggregation of large-scale models of  ecosystems based on ODEs or difference equations (the time-discrete homologous of ODEs).  
Aggregations are obtained as a coarsening of the state space in terms of \emph{macro-variables}, each describing a group of original variables (i.e., of entities in the original ecosystem). Aggregations have been developed to ignore, e.g., age differences in population models~\cite{Iwasa1987287} or to exploit the fact that entities act at different time scales (see~\cite{Auger2008} for a survey).

\item
Cellular automata are a classic example of models used in physics to describe dynamical systems consisting of simple agents with small state spaces, whose interactions result in complex phenomena. Abstraction methods identify sets of neighbouring cells for which a coarse-grained description can still describe the overall behaviour of the original model~\cite{PhysRevE.73.026203}.

\item
In computational systems biology, many abstraction techniques have been developed to cope with the combinatorial species and chemical reaction networks (CRN) that arise from the interactions of basic molecules which can undergo internal changes, or that can bind each other to form complex species. 
In particular, we refer to techniques developed for systems of ODEs arising from protein interaction networks (e.g.~\cite{Borisov2005951,Feret21042009}). Here, 
methods have been proposed to consider:  a covering of the state space (where a variable may appear in more than one group)~\cite{Feret21042009}; quotienting, induced by a partition of the ODE variables (e.g.,~\cite{Feret2012137}); and aggregations exploiting time-scale separation (e.g.,~\cite{Murray2002}).

\item
Large-scale dynamical systems are common also in control engineering. Here, starting from the seminal paper~\cite{1098900}, many approaches have been proposed to reduce the original model in a way that preserves  controllability, i.e., the capability of driving the system to a desired state by using appropriate control inputs~\cite{Pappas}. 
\end{itemize}

More recently, in the computer science community there has been an increasing interest towards quantitative models of computation based on ODEs, for example to use formal languages to describe biochemical models~\cite{1999133,Danos200469,Blinov22112004,Calzone15072006,CardelliTCS08,DBLP:conf/cmsb/KwiatkowskiS08,biopepa,Pedersen:2010aa} 
or as a deterministic approximation for languages with stochastic semantics~\cite{biopepa,DBLP:journals/tcs/HaydenB10,TribastoneTSE2012}.


Our own line of research~\cite{tacas17cttv,gecco17,DBLP:conf/lics/CardelliTTV16,DBLP:conf/popl/CardelliTTV16,DBLP:conf/sfm/VandinT16,DBLP:conf/tacas/CardelliTTV16,DBLP:conf/concur/CardelliTTV15,DBLP:conf/mfcs/IacobelliTV15} consists of a computer-science perspective  to the abstraction problem, borrowing ideas from the concurrency theory community. 
We recast the ODE reduction problem to that of finding an appropriate equivalence relation over ODE variables, akin to classical models of computation based on labelled transition systems.
We studied such \emph{differential equivalences} for two basic intermediate languages, trading expressivity for efficiency: 
\begin{enumerate}
\item[i)] \emph{IDOL} (Intermediate Drift-Oriented Language)~\cite{DBLP:conf/popl/CardelliTTV16} covers a general class of non-linear ODEs with derivatives containing polynomials, rationals, minima/maxima, and absolute values. This is enough, e.g., to capture the existing ODE semantics of stochastic process algebras~\cite{DBLP:conf/mfcs/IacobelliTV15,Tschaikowski2013b,concur2012,CardelliTCS08}.  
The largest equivalences of IDOL terms are computed using a symbolic partition-refinement algorithm that exploits an encoding into a satisfiability modulo theories (SMT) problem;
\item[ii)]
\emph{Reaction networks}~\cite{DBLP:conf/concur/CardelliTTV15,DBLP:conf/tacas/CardelliTTV16}, a slight generalization of chemical reaction networks,  characterise ODEs with polynomial derivatives. 
In this case, the partition refinement is based on Paige and Tarjan's seminal proposal~\cite{partitionref}, giving an efficient algorithm that runs in polynomial time. 
\end{enumerate}

Our framework for ODE reduction has been implemented in the tool ERODE~\cite{tacas17cttv} (\url{http://sysma.imtlucca.it/tools/erode/}), allowing us to provide evidence of effective reductions in realistic models from the literature. 

This paper briefly reviews our framework for ODE reduction. 
A more detailed tutorial-like presentation unifying the two approaches can be found in~\cite{DBLP:conf/sfm/VandinT16}, while~\cite{DBLP:conf/lics/CardelliTTV16,gecco17} address the more general problem of computing all differential equivalences of a model.

\section{Framework Overview}

\paragraph{Differential equivalences.}
Differential equivalences induce a quotienting of the ODE variables. Two distinct notions of equivalence for ODEs have been provided in~\cite{DBLP:conf/popl/CardelliTTV16}. \emph{Forward differential equivalence} (FDE) allows us to abstract from single variables, and to consider only cumulative information on each equivalence class. In particular, it guarantees that the original ODE system can be described in terms of a smaller ODE system having only one macro-variable per equivalence class that describes the sum of the variables within the equivalence class. For instance, consider the following ODE system with variables $x_1$, $x_2$, and $x_3$: 
\begin{align}\label{eq:simple}
\dot{x}_1 & = - x_1, &  \dot{x}_2 & = k_1 \cdot x_1 - x_2, & \dot{x}_3 & = k_2 \cdot x_1 - x_3,
\end{align}
where $k_1$ and $k_2$ are constants and the ‘dot' operator denotes the derivative operator (with respect to time).
It can be shown that the partition of variables $\pa=\{\{x_1\},\{x_2,x_3\}\}$ is an FDE because
\begin{align}\label{eq:simple.fde}
\dot{x}_1 & = - x_1,  & \dot{(x_2 + x_3)} & = \dot{x}_2 + \dot{x}_3 = (k_1 + k_2) \cdot x_1 - (x_2 + x_3) .
\end{align}
In fact, by applying the simple change of variable $x_{2,3} = x_2 + x_3$, this is equivalent to writing
\begin{align*}
\dot{x}_1 & = - x_1 & \dot{x}_{2,3} & = (k_1 + k_2) \cdot x_1 - x_{2,3} .
\end{align*}
This reduced ODE system contains one variable per equivalence class in $\pa$, each of which describes the sum of the solutions of the variables in the corresponding equivalence class. Thus, setting the \emph{initial condition} $x_{2,3}(0) = x_2(0) + x_3(0)$ yields that the solution satisfies $x_{2,3}(t) = x_2(t) + x_3(t)$ at all time points $t$. 
As discussed in~\cite{DBLP:conf/concur/CardelliTTV15}, the notion of FDE has proved to be particularly useful in the context of computational systems biology, where modellers are interested in a few observations of interest, like  the evolution of the concentration of few species only, or of the sums of certain species that represent different configurations of the same molecule.

\emph{Backward differential equivalence} (BDE) instead allows us to identify redundant dynamics in the original system. In fact, it equates variables that have the same solutions at all time points, if initialized equally. In the ODE system in Equation~\eqref{eq:simple}, if $k_1=k_2$, we have that $\pa=\{\{x_1\},\{x_2,x_3\}\}$ is also a BDE. Hence, we can reduce the original ODE system by removing either equation between $x_2$ and $x_3$, say $x_3$, and by rewriting every remaining occurrence of $x_3$ as $x_2$:
\begin{align*}
\dot{x}_1 & = - x_1 & \dot{x}_2 & = k_1  x_1 - x_2 .
\end{align*}
The notion of BDE has proved to be useful in the context of evolutionary biology, where one is interested in understanding whether a system has evolved into another one preserving the original functionality~\cite{cardelli14,ccdtt-sci-reports-2012,10.1371/journal.pcbi.1005100}. 
In fact, it is well known that using larger networks instead of smaller ones of \emph{equal functionality} is beneficial for enhanced stability with respect to stochastic noise~\cite{ccdtt-sci-reports-2012}. 
In particular, in~\cite{DBLP:conf/popl/CardelliTTV16,DBLP:conf/lics/CardelliTTV16} we have shown that BDE succeeded in tracing a certain functionality across two CRNs (by taking the CRN whose species and reactions are the disjoint union of those from the source and target one).

 Differential equivalences guarantee that the relationship between the original model and the abstract one is exact. However, FDE leads to a loss of information, because information on the individual variables within an equivalence class may not be recovered in general.
Instead, BDE preserves all information, however it can be applied only when the initial conditions are coherent with the considered equivalence classes: i.e., when all backward equivalent variables are initialized equally. 
Forward differential equivalence is closely related to the notion of exact ODE lumpability, thoroughly investigated in the chemistry domain (e.g.,~\cite{LiRabitz1997,okino1998,Li19891413}). However no automated procedure exists to reduce an ODE system using this approach (e.g.,~\cite{Turanyi14}). To cope with this, i.e., to guarantee that our framework can be instantiated in a family of fully automatic reduction techniques, restrictions are imposed to be able to develop minimisation algorithms.

\paragraph{Symbolic minimisation.}
In~\cite{DBLP:conf/popl/CardelliTTV16} each ODE variable is treated explicitly as a real function and a differential equivalence is encoded in a logical formula over ODE variables. Thus, checking whether a candidate partition is BDE/FDE can be done symbolically using an encoding into satisfiability modulo theories (SMT)~\cite{Biere:2009:HSV:1550723}. In fact, differential equivalences belong to the quantifier-free fragment of first-order logic. It is possible to restrict the admissible ODE systems to those for which an SMT solver for nonlinear real arithmetic — e.g.~Z3, \cite{DeMoura:2008:ZES:1792734.1792766} — is a decision procedure. This can be done by, roughly speaking, excluding trigonometric functions. Hence, e.g., we support rational functions, necessary to encode sophisticated biological kinetics like Hill's one~\cite{Biere:2009:HSV:1550723}, and minima/maxima, necessary to capture the ODEs arising from terms of stochastic process algebras~\cite{DBLP:journals/tcs/HaydenB10,TribastoneTSE2012,cj:fluidflow}, or from queuing networks~\cite{10.1109/TSE.2012.66}.
Hence, the technique is generic enough to be applied to a wide range of dynamical systems. 

Let us consider the example in Equation~\eqref{eq:simple}, assuming $k_1 = k_2 = 1$, for which case we have shown that $\pa=\{\{x_1\},\{x_2,x_3\}\}$ is a BDE. The condition for $\pa$ to be a BDE has been shown to correspond to requiring that related variables with equal assignments always have equal derivatives~\cite{DBLP:conf/popl/CardelliTTV16}. This can be encoded in a quantifier-free first-order logical formula, which in the specific case of  $\pa$ is:
\[
\phi_{\pa} \ := \ (x_2 = x_3) \implies (k_1\cdot x_1 -x_2 =k_2\cdot x_1 - x_3)
\]
The SMT check $sat(\neg\phi_\pa)$ searches for an assignment of the variables $x_1$, $x_2$, and $x_3$ such that $\neg\phi$ holds. Hence, the partition is a BDE if and only if the procedure returns ``unsat'', meaning that $\phi_\pa$ can not be falsified. This is the case for our example. More interestingly, in case the solver returns an assignment which satisfies $\neg\phi_\pa$, a \emph{witness}, then we can use it within an iterative algorithm which computes the coarsest BDE partition that refines the initial one (i.e., that can be obtained by splitting the blocks of the initial one). 
%
%
Assume for example that we start from the candidate BDE partition $\pa_2=\{\{x_1,x_2,x_3\}\}$. It can be shown that $\pa_2$ is not a BDE, and hence the SMT check $sat(\neg\phi_{\pa_2})$ is satisfiable. For instance, a witness for the satisfiability of $\neg\phi_{\pa_2}$ is $(x_1 = 1, x_2 = 1, x_3 = 1)$, which yields to different evaluations for the three derivatives: 
\[
\dot{x}_1 = -x_1= −1
,\quad
\dot{x}_2 = k_1\cdot x_1-x_2 =0
, \quad
\dot{x}_3 = k_2\cdot x_1 - x_3 = 0
\] 
This suggests the implementation of a partition-refinement algorithm that \emph{splits} the blocks of the current partition in sub-blocks of variables which cannot be distinguished using the returned witness. In other words, at the next iteration the candidate BDE partition would be $\{\{x_1\},\{x_2,x_3\}\}$. 
%
A similar algorithm has been provided also for the FDE case. 
Note that this algorithm meets an important property often required for reduction techniques: the reduced model preserves user-defined observables of the original system.
 Indeed, each variable of interest for the modeller can be put in a singleton initial block.  In the backward case, this allows us to provide an initial partition coherent with the initial conditions of the original model (that is, two variables are in the same initial block if their initial conditions are the same), a necessary condition for the reduced model to faithfully represent the original dynamics. 
 %

\paragraph{Syntactic minimisation.}
More efficient partition refinement  algorithms can be provided for ODEs with derivatives that contain only multivariate polynomials of degree at most two~\cite{DBLP:conf/tacas/CardelliTTV16}. This is quite a general class of ODEs, covering linear systems as well as chemical reaction networks. This approach is based on 
a syntactic representation of an ODE system in terms of a so-called \emph{reaction network} (RN), which slightly generalizes the notion of chemical reaction network by allowing reaction rates to be also negative. Therefore, one can readily encode any polynomial ODE system as an RN~\cite{DBLP:conf/tacas/CardelliTTV16}. In particular, an RN consists of a set of species/variables interacting by means of reactions parameterised by a real value. 
Intuitively, we use reaction networks to generalize labelled transition systems (LTS) so to consider transitions among multisets of nodes rather than among individual nodes. Hence, there exists a loose connection between species and nodes of an LTS, and between reactions and transitions of an LTS. 

We provide two bisimulation equivalences for such syntactic representation:  the forward RN bisimulation (FB), and the backward RN bisimulation (BB), which are related to FDE and BDE, respectively~\cite{DBLP:conf/concur/CardelliTTV15}. Continuing the parallel between reaction networks and LTSs, our bisimulations are similar in spirit to quantitative bisimulations on LTSs, e.g., Larsen and Skou's probabilistic bisimulation~\cite{Larsen19911}. In particular, the largest RN bisimulations that refine a given input partition can be computed by generalizing Paige and Tarjan's famous algorithm~\cite{partitionref}. In~\cite{DBLP:conf/tacas/CardelliTTV16} we propose a partition refinement algorithm along the lines of the best-performing analogues for Markov chain lumping such as~\cite{Derisavi2003309} and~\cite{DBLP:conf/tacas/ValmariF10}, and for probabilistic transition systems~\cite{DBLP:journals/jcss/BaierEM00}. The algorithm computes the largest FB/BB refining a given input partition of variables in $O(mn\log n)$ time, where $m$ is the number of monomials in the ODE system and $n$ is the number of variables.

\paragraph{Tool support: ERODE.}
We implemented our minimisation techniques in ERODE~\cite{tacas17cttv}, a mature tool featuring a modern integrated development environment for the evaluation and reduction of ordinary differential equations. 
%
ERODE is on the Eclipse framework, it is multi-platform and does not require any installation process. The tool is available together with a manual and sample models at \url{http://sysma.imtlucca.it/tools/erode}.

\begin{figure}[t]
\centering
\includegraphics[width=0.9\textwidth]{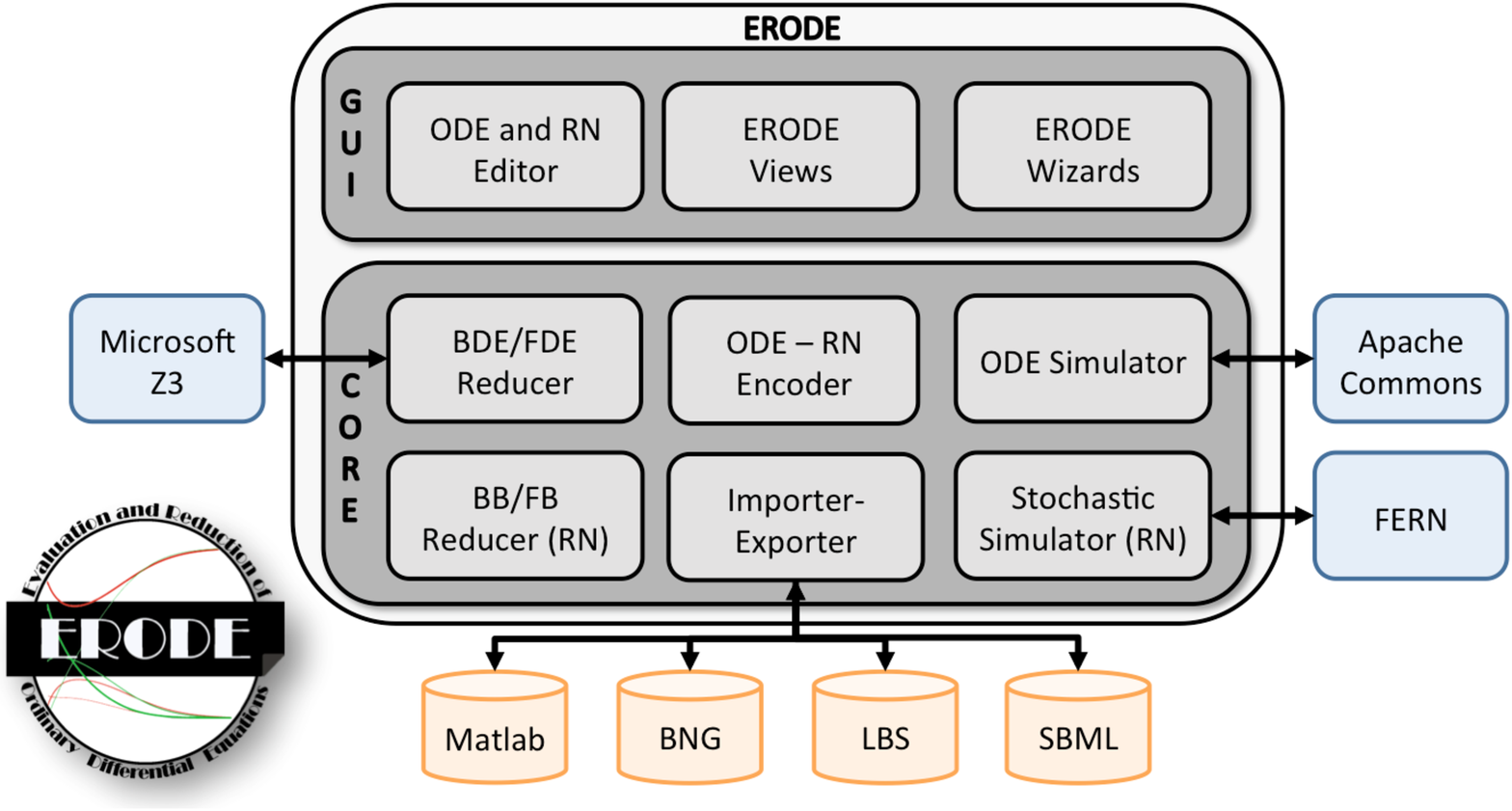}
\caption{\erode{}'s Architecture.}\label{fig:architecture}
\end{figure}

\begin{figure}[t]
 \centering
\includegraphics[width=1.0\textwidth]{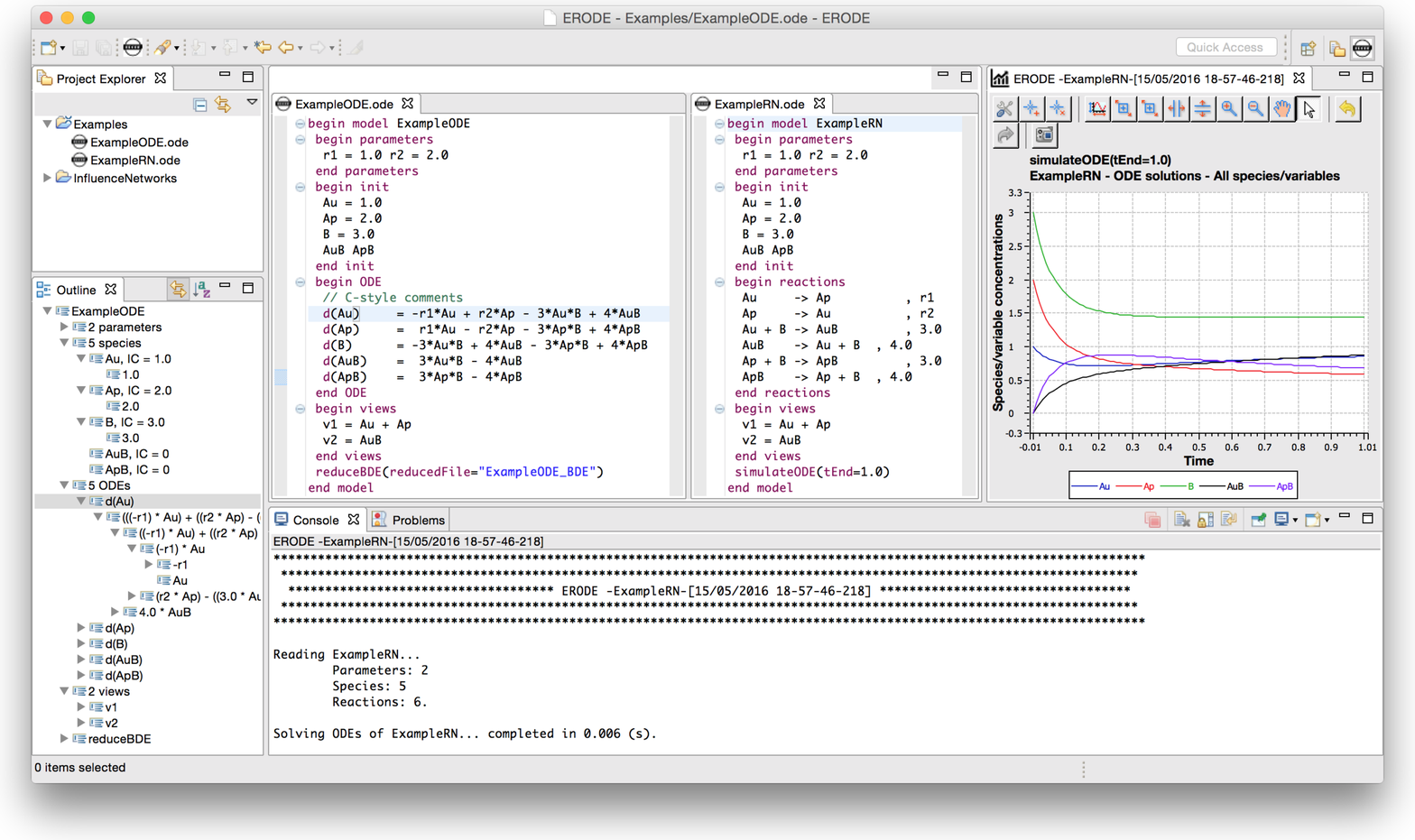}
 \caption{A screenshot of \erode{}. 
 }\label{fig:screenshot}
 \end{figure}

Fig.~\ref{fig:architecture} depicts 
\erode{}'s architecture. 
It is organized in the GUI layer, and the core layer.
Fig.~\ref{fig:screenshot} depicts the main components of the GUI layer, including 
 a fully-featured text editor based on the \textsl{XTEXT} framework which supports syntax highlighting, content assist, error detection,  and fix suggestions (top-middle of Fig.~\ref{fig:screenshot}).
Additionally, this layer offers a number of views, including a \emph{project explorer} to navigate among different \erode{} files (top-left of Fig.~\ref{fig:screenshot});
an \emph{outline} to navigate the parts of the currently open \erode{} file (bottom-left of Fig.~\ref{fig:screenshot});
a \emph{plot view} to display ODE solutions (top-right of Fig.~\ref{fig:screenshot}); and a \emph{console view} to display diagnostic information (bottom-right of Fig.~\ref{fig:screenshot}). 
%

The core layer implements the minimization algorithms and related data structures for FDE, BDE, FB and BB. 
The algorithms for FB and BB reductions have been provided entirely in Java, while those for FDE and BDE reductions resort on the Z3 SMT solver, accessed via its Java APIs. 
The core layer also features encoding capabilities from RN to ODE (or IDOL) representation, and vice versa, as well as  export/import functionalities for third-party formats, like biochemical models for 
BioNetGen~\cite{Blinov22112004} and Microsoft GEC~\cite{GEC}, or ODEs defined in MATLAB. Also, models can be exported in the SBML format, an XML-based language for describing biological systems. 
Finally, 
this layer provides support for the numerical solutions of ODEs by means of  the Apache Commons Maths library~\cite{apache:maths}. 
When the input is a CRN (i.e. an RN with only positive rates) it can also be interpreted as a CTMC, following an established approach~\cite{Gillespie77}. Using the \emph{FERN} library~\cite{Erhard2008}, \erode{} features CTMC simulation. 

By using a standard laptop machine, in~\cite{tacas17cttv} we have shown that our syntactic minimisation algorithms can scale to models with up to 2.5 million variables and 25 million monomials, with runtimes of a few minutes only. Instead, the symbolic minimisation scales up to 70 thousand variables in the backward case, and to 1 thousand  variables in the forward case. 
As discussed in~\cite{DBLP:conf/popl/CardelliTTV16}, the difference in performance and scalability between the backward and forward symbolic minimisation techniques are due to the fact that combinatorially many SMT checks among each pair of related variables are required in order to establish if a partition is FDE, while the BDE case requires only one check.
%

\section{Conclusions}

This paper presented two equivalences for ordinary differential equations (ODEs), forward and backward differential equivalences. 
In case the ODEs can be described in terms of the IDOL language, which essentially restricts to ODEs whose derivates do not contain exponential and trigonometric functions, then such ODE systems can be minimized up to these notions of equivalence using symbolic reduction techniques based on SMT solving. 

If instead the derivatives of the considered ODE system are multivariate polynomials, then more efficient syntactic minimization techniques can be used, based on an encoding of the notions of differential equivalence in terms of two probabilistic bisimulations, forward and backward RN bisimulations. 
In~\cite{tacas17cttv} we have shown that the syntactic minimisation algorithms can scale to systems with millions of variables and monomials, terminating in a few minutes also in some challenging models. There are, however, further challenges ahead which we wish to tackle.

Forward bisimulation is only a sufficient condition for FDE; while it provides significant reductions in practice~\cite{DBLP:conf/concur/CardelliTTV15,DBLP:conf/tacas/CardelliTTV16}, some other examples from the literature demonstrate that the algorithm may miss some FDE reductions (see~\cite{DBLP:conf/popl/CardelliTTV16}). In ongoing work we are relaxing the notion of forward bisimulation so to fully characterize FDE 
(for multivariate polynomial derivatives of degree at most two). Instead, backward bisimulation characterises BDE, and hence the corresponding syntactic minimisation technique from~\cite{DBLP:conf/tacas/CardelliTTV16} should be used when dealing with polynomial ODEs of degree at most two. Also, we plan to extend the RN representation and the two notions of forward and backward RN bisimulations in order to be able to apply them ti to polynomial derivatives with arbitrary degree.  

As regards our symbolic reduction techniques, we used SMT only in a black box fashion. It could be interesting to study ad-hoc heuristics for SMT solving that could lead to better performance or scalability. For the same reason, it would be interesting to consider parallelised versions of such symbolic reduction techniques.   
Also, another interesting research line related to our symbolic reduction techniques is that of further exploiting the generality of SMT in order to handle models with uncertainties in rates (a well-known issue in mathematical biology): here the SMT framework can already be easily extended to compute partitions that are differential equivalences under all possible assignments of such uncertain parameters, left as free variables in the satisfiability problem (similarly to the SMT-based parametric minimisation approach of~\cite{DKP13} for probabilistic models written in PRISM~\cite{DBLP:conf/cav/KwiatkowskaNP11}).

Lastly, we remark that our techniques regard exact aggregations. In some cases, however, one might be interested in more permissive, approximate notions that do not discriminate ODE variables with nearby trajectories 
%
(e.g.,~\cite{DBLP:conf/concur/PierroHW03,worrell:approximating,DBLP:journals/lmcs/GuptaJP06,tac15,dsn2013,DBLP:journals/tcs/DesharnaisGJP04,DBLP:conf/mfcs/LarsenMP12}). 
Such approximate variants can be considered as weaker notions that, e.g., allow variability in the parameters, considering the exact versions as a degenerate case in which no such variability is needed.
We remark that, at least for the numerical benchmarks considered so far with realistic models, the exact reductions can already be quite effective. Of course, approximate ones might be able to provide even coarser descriptions. In this case, however, the main challenge is to be able to relate the variability in the parameters tolerated by the coarsening procedure with the error incurred when considering an approximate, smaller model, instead of the original one.
In an ongoing research we are developing approximate variants of our differential equivalences, aiming at maintaining computational tractability, and certified error bounds that do not grow fast with time. 

As regards our tool \erode{}, we also plan to enrich its family of offered analysis techniques. For example, we plan to extend \erode{}, and in particular its stochastic simulator, with statistical model checking~\cite{legay2010statistical} capabilities. %
This can be obtained upon integration with the statistical analyser MultiVeStA~\cite{sebastio2013multivesta}, which has already proven useful in the analysis of a wide range of scenarios, including software product lines~\cite{DBLP:conf/splc/BeekLLV15,DBLP:conf/isola/BeekLLV16}, crowd-steering~\cite{DBLP:conf/hpcs/PianiniSV14}, 
public transportation systems~\cite{DBLP:conf/ifm/GilmoreTV14},  
and swarm robotics~\cite{MisscelAndPirlo}.

\paragraph{Acknowledgement.}
The author is indebted to Luca Cardelli, Mirco Tribastone, and Max Tschaikowski for the collaboration on the development of the presented framework, and for many fruitful 
discussions. 

\bibliographystyle{eptcs}
\bibliography{qapl2017}
\end{document}